%% file: tabby_asassn_v4.tex
\documentclass[iop]{emulateapj}

\usepackage{color}
\usepackage[dvipsnames,svgnames]{xcolor}

\usepackage{natbib}

\renewcommand{\dh}{\fontencoding{T1}\selectfont{\symbol{240}}} 

\def\spose#1{\hbox to 0pt{#1\hss}}
\def\lta{\mathrel{\spose{\lower 3pt\hbox{$\mathchar"218$}}
     \raise 2.0pt\hbox{$\mathchar"13C$}}}
\def\gta{\mathrel{\spose{\lower 3pt\hbox{$\mathchar"218$}}
    \raise 2.0pt\hbox{$\mathchar"13E$}}}

\shorttitle{The Long-Term Variability of Boyajian's Star}
\shortauthors{Simon et al.}
\slugcomment{ApJ, in press}

\begin{document}

\title{Where Is the Flux Going?  The Long-Term Photometric Variability
  of Boyajian's Star}

\author{Joshua D. Simon\altaffilmark{1}, 
  Benjamin J. Shappee\altaffilmark{1,2,3},
  G.~Pojma\'nski\altaffilmark{4},
  Benjamin T. Montet\altaffilmark{5,6},
  C.~S. Kochanek\altaffilmark{7,8},
  Jennifer van Saders\altaffilmark{1,3},
  T.~W.-S. Holoien\altaffilmark{7,8,9},
  and Arne A. Henden\altaffilmark{10}
}

\altaffiltext{1}{Observatories of the Carnegie Institution for
  Science, 813 Santa Barbara St., Pasadena, CA 91101;
  jsimon@carnegiescience.edu}

\altaffiltext{2}{Hubble Fellow}

\altaffiltext{3}{Carnegie-Princeton Fellow}

\altaffiltext{4}{Warsaw University Observatory, Al. Ujazdowskie 4, 00-478 Warsaw, Poland}

\altaffiltext{5}{Department of Astronomy and Astrophysics, University of Chicago, 5640 S. Ellis Avenue, Chicago, IL 60637}

\altaffiltext{6}{NASA Sagan Fellow}

\altaffiltext{7}{Department of Astronomy, The Ohio State University, 140 West 18$^{th}$ Avenue,
  Columbus, OH 43210}

\altaffiltext{8}{Center for Cosmology and Astroparticle Physics, The Ohio State University,
  191 W. Woodruff Avenue, Columbus, OH 43210}

\altaffiltext{9}{US Department of Energy Computational Science Graduate Fellow}

\altaffiltext{10}{AAVSO, 49 Bay State Road, Cambridge, MA  02138}

\begin{abstract}
We present $\sim800$ days of photometric monitoring of Boyajian's Star
(KIC~8462852) from the All-Sky Automated Survey for Supernovae
(ASAS-SN) and $\sim4000$ days of monitoring from the All Sky Automated
Survey (ASAS).  We show that from 2015 to the present the brightness
of Boyajian's Star has steadily decreased at a rate of $6.3 \pm
1.4$~mmag~yr$^{-1}$, such that the star is now 1.5\%\ fainter than
it was in February 2015.  Moreover, the longer time baseline afforded 
by ASAS suggests that Boyajian's Star has also undergone two 
brightening episodes in the past 11 years, rather than only exhibiting 
a monotonic decline.  We analyze a sample of $\sim1000$ comparison
stars of similar brightness located in the same ASAS-SN field and
demonstrate that the recent fading is significant at
$\gtrsim99.4$\%\ confidence.  The $2015-2017$ dimming rate is 
consistent with that measured with \emph{Kepler} data for the time 
period from 2009 to 2013.  This long-term variability is difficult 
to explain with any of the physical models for the star's behavior 
proposed to date.
\end{abstract}


\section{INTRODUCTION}
\label{intro}

Boyajian's Star (KIC~8462852) has been a fascinating mystery since its
discovery was made public in late 2015.  \citet{boyajian16} showed
that the star underwent a series of asymmetric, aperiodic, brief dips
in brightness from 2009 to 2013 while it was being monitored by
\emph{Kepler}.  These dips have typical timescales of a few days and
range in depth from $\sim0.2$\%\ to more than 20\%.
\citet{boyajian16} cataloged ten dip complexes detected during the
\emph{Kepler} mission.

Such irregular dimming events are not unusual for young stars
\citep[e.g.,][]{alencar10,morales-calderon11,cody14}, but Boyajian's
Star exhibits no obvious signs of youth \citep{boyajian16}.  Instead,
it appears to be an otherwise unremarkable main sequence F star
\citep{lisse15}.  The star also lacks an infrared excess that would be
expected to accompany significant amounts of nearby warm dust
\citep{marengo15,thompson16}.

Initial attempts to explain the dips focused on circumstellar
material, perhaps in the form of debris from a collision of
planetesimals \citep{boyajian16} or a large family of comets
\citep{bq16}.  While these models can potentially account for the
dipping activity without violating the infrared constraints, they were
severely challenged by the finding by \citet{schaefer16} that
Boyajian's Star appears to have been monotonically fading since the
late 19th century, with a total change in its brightness of 0.16~mag
from 1890 to 1989.  The interpretation of the archival photographic
measurements for the star has been controversial
\citep{hippke16,hippke17}, but \citet{ms16} demonstrated using
\emph{Kepler} full-frame images that the star's brightness did
steadily decline by a total of $\sim3$\%\ over the four years of
\emph{Kepler}'s main mission.  Further complicating the situation,
\citet{mg16} used astrometry from \emph{Kepler} to show that the
variability detected by \citet{boyajian16} in the \emph{Kepler}
long-cadence light curve must originate from at least two distinct
sources.  While the strong dips can only occur on Boyajian's Star
given its brightness relative to the closest neighboring stars, the
\citeauthor{mg16} results indicate that the longer-term, low-amplitude
periodicities (ranging from 0.88~d to $\sim20$~d) are most likely
attributable to a nearby star or stars.

After the long-term variability of Boyajian's Star became clear,
\citet{ws16} provided a comprehensive summary of all possible
mechanisms to account for the observed brightness changes, and
\citet{metzger17} and \citet{foukal17} presented new models for
possible intrinsic variability of the star.  In particular,
\citet{metzger17} show that the ingestion of a planet and the
accompanying destruction of its satellite system could produce both
brief dips and a long-term dimming with a rate comparable to that
measured by \citet{schaefer16} and \citet{ms16}, providing the first
simultaneous explanation for all aspects of the star's variability.
\citet{foukal17} argues instead that the convective structure of the
star may be able to store enough energy to explain the dips and
longer-term fading.

In this paper, we examine the long-term photometric behavior of
Boyajian's Star from the ground using data from the All-Sky Automated
Survey for Supernovae \citep[ASAS-SN;][]{shappee14} and the All Sky
Automated Survey \citep[ASAS;][]{asas}.  In \S\ref{observations} we
describe the ASAS-SN and ASAS imaging of the \emph{Kepler} field.  In
\S\ref{analysis} we analyze the ASAS-SN observations of Boyajian's
Star and our comparison sample.  In \S\ref{asas} we analyze the ASAS
observations.  We discuss the interpretation of these results in light
of ideas suggested to explain the behavior of Boyajian's Star in
\S\ref{discussion} and we present our conclusions in
\S\ref{conclusions}.

\section{OBSERVATIONS}
\label{observations}

\subsection{ASAS-SN}

ASAS-SN is an all-sky survey that monitors the entire sky for
transients every $\sim 2$ days down to a $V$-band magnitude of $\sim
17$. The survey currently consists of two fully robotic units, one on
Haleakala, Hawaii and one on Cerro Tololo, Chile, each with four 14~cm
telescopes.  The field containing Boyajian's Star was observed a total
of 377 times from 2015 February 24 (UT) to 2017 May 15 by the northern
ASAS-SN unit, ``Brutus".  This ASAS-SN field is labeled
F2006+46\_0982, which is centered on R.A.$ = 20$:06:46.5, Decl.$ =
+45$:59:27.8 and is observed with the Brutus camera labeled ``bd''.
In order to provide a comparison sample of stars with similar
brightness, we used the SIMBAD database \citep{simbad} to select
$\sim1100$ stars located in the same ASAS-SN field and within
$\sim0.5$~mag of the brightness of Boyajian's Star ($11.5 < V <
12.5$).

We performed aperture photometry for each of these stars using
automated measurements similar to those described by
\citet{kochanek17} except that we align, interpolate, and stack the
dithered images (usually 3) acquired during each epoch.
Experimentation suggested that the most robust results for our
purposes could be obtained using a larger photometric aperture of 3
pixels ($\sim24$\arcsec).  The median photometric uncertainty for
these choices is 0.010~mag.  The photometric calibration for each star
is determined from the magnitudes of $11.5 < V < 13.5$ stars in the
AAVSO All-Sky Photometric Survey \citep[APASS;][]{Henden12} catalog
located within 0\fdg75 that also have photometric uncertainties less
than 0.1~mag.  Using fainter or more distant APASS stars or larger
photometric apertures tended to produce poorer results for this field
and brightness range.

\subsection{ASAS}

ASAS is a long-term variability survey that scans a large fraction of
the sky from two locations, Las Campanas, Chile (since 1997) and
Haleakala, Hawaii (since 2006).  From Hawaii, ASAS simultaneously
monitors the sky in two filters ($V$ and $I$) with two 10~cm
telescopes.  ASAS $V$-band aperture photometry is calibrated to Tycho
$V$ magnitudes by determining the median difference between ASAS
instrumental magnitudes and Tycho in $30\arcmin \times 30\arcmin$
regions.  The zero point for each $8\degr \times 8\degr$ ASAS field is
then calculated by interpolating the results from the $16 \times 16$
grid of regions with a spatially smooth function.  ASAS has obtained a
total of 564 $V$-band images of Boyajian's Star spanning from 2006 May
31 to the beginning of the 2017 May dipping activity.  A large
majority of these images (517) are in ASAS field 616, centered at
R.A.$ = 20$:15:00, Decl.$ = +48$:00:00.  As with the ASAS-SN data, we
consider only observations in the same field in order to minimize
systematics between Boyajian's Star and comparison objects.  Relative
to ASAS-SN, ASAS has the advantage of a much longer time baseline, but
the disadvantage of larger photometric uncertainties, with a median
uncertainty of 0.028~mag per measurement.

\section{ANALYSIS OF ASAS-SN DATA}
\label{analysis}

\subsection{Boyajian's Star}
\label{tabbysstar}

\begin{figure*}[th!]
\epsscale{1.0}
\plotone{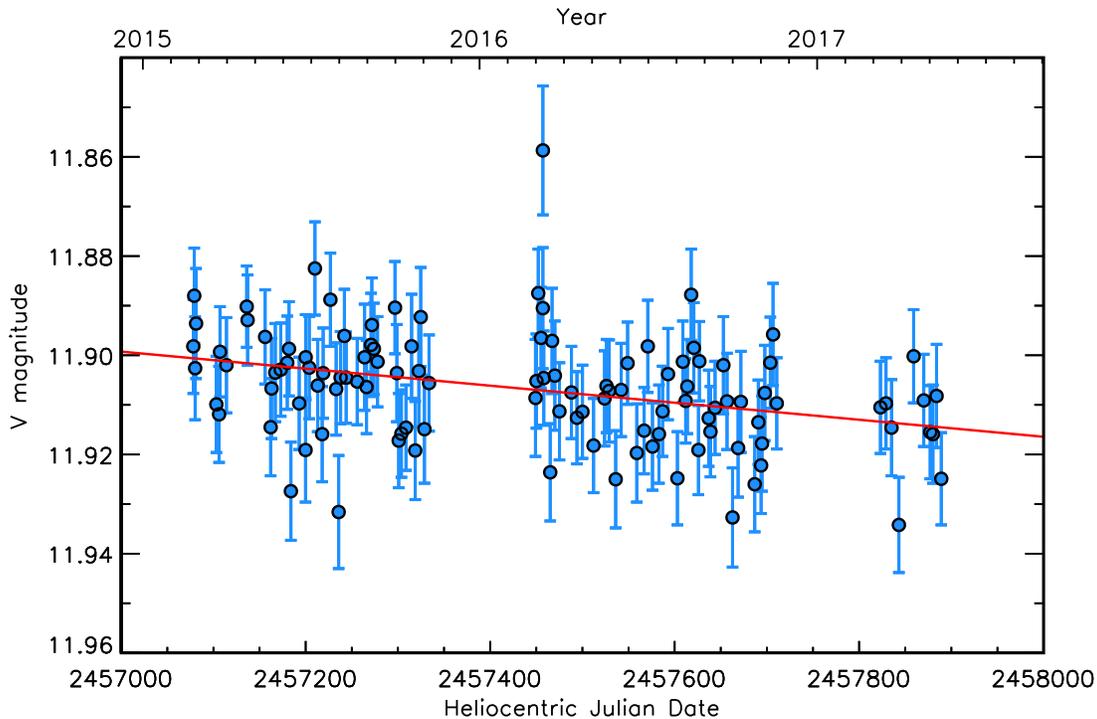}
\caption{ASAS-SN light curve of Boyajian's Star.  The red line is a
  least squares fit to the data showing a decline in brightness of
  $6.3 \pm 1.4$~mmag~yr$^{-1}$.}
\label{tabby_lc}
\end{figure*}

The ASAS-SN light curve of Boyajian's Star is displayed in
Figure~\ref{tabby_lc}, and the data are listed in
Table~\ref{asassn_data_table}.  The measurements reveal a small but
clear decline in brightness from $V = 11.899$ in early 2015 to $V =
11.914$ in 2017 May.  While ASAS-SN observations of the star will
continue with a few-day cadence until it goes into conjunction with
the Sun in 2017 November, we exclude measurements after 2017 May 16 to
ensure that our results are not affected by the ``Elsie'' dip
beginning on that date \citep{boyajian17}.  We also remove images that
were acquired in adverse conditions such as bright sky, low
transparency, or poor seeing by rejecting measurements when the
photometric uncertainty was $>0.03$~mag, the transparency was more
than 0.5~mag worse than the median value, the image FWHM was more than
$1\sigma$ above the median value, or the $V$ magnitude was $>0.1$~mag
away from the median magnitude.  We perform an ordinary least squares
(OLS) fit to the data assuming that the magnitude of the star varies
linearly with time and find a best-fit slope of $6.3 \pm
1.4$~mmag~yr$^{-1}$.  The change in brightness is detected at
$4.4\sigma$.  The scatter of the data points around the fit is
0.011~mag and the $\chi^{2}$ per degree of freedom is 1.08.  If we
measure the flux of Boyajian's Star in apertures with radii of 2, 4,
or 5 ASAS-SN pixels, we find that the best-fit slope changes by less
than its uncertainty, and the significance varies from $5.5\sigma$ at
$r=2$~pixels to $3.5\sigma$ at $r=4$~pixels.

\input{asassn_data_table_stub.tex}

While the assumptions of the OLS method are violated by many
astronomical data sets \citep[e.g.,][]{ab96,kelly07,hogg10}, this
particular case is relatively well-suited for OLS: the times at which
the magnitudes were measured are known essentially perfectly and the
uncertainties on the magnitudes are small (and only mildly
heteroscedastic).  We do not know a priori whether a linear model is a
good fit to the brightness of Boyajian's Star over this time interval,
but it is apparent from Fig.~\ref{tabby_lc} that the model is able to
describe the data and more complicated functional forms are not
justified.  That the $\chi^{2}$ per degree of freedom is somewhat
larger than unity suggests that the photometric errors are slightly
underestimated or that there is intrinsic scatter in the data.  To
explore this point, we fit the data with the linmix\_err code from
\citet{kelly07}.  We find an intrinsic scatter of 0.0035~mag, which is
significantly smaller than the measurement uncertainties.  The
resulting slope of $6.3 \pm 1.5$~mmag~yr$^{-1}$ is consistent with the
OLS fit but has slightly larger uncertainties such that the trend of
magnitude with time is non-zero at the $4.1\sigma$ level.

Over the four-year \emph{Kepler} main mission \citet{ms16} detected a
decrease in the brightness of Boyajian's Star of 8.4~mmag~yr$^{-1}$,
in agreement with the rate of fading measured here given the
uncertainties.  In the \emph{Kepler} data, though, the decline was not
consistently linear with time; \citeauthor{ms16} observed fading of
$3.7 \pm 0.4$~mmag~yr$^{-1}$ for the first 2.7~yr, followed by a rapid
drop of nearly 2.5\%\ over the next $\sim6$~months.  With the shorter
time span of the ASAS-SN data a purely linear decrease is plausible,
and modest departures from linearity would likely not be detectable
given the photometric uncertainties.  We also note that the
\emph{Kepler} bandpass \citep{Koch10} is broader and redder than the
ASAS-SN $V$ filter.  How this difference affects the relative decline
rates depends on whether the fading of Boyajian's Star is chromatic or
achromatic.

\subsection{Comparison Samples}

Although the change in brightness of Boyajian's Star in the ASAS-SN
data is statistically significant given the assumed errors and the
observed scatter, ASAS-SN has not previously been used to study very
low amplitude variability over timescales of years.  We therefore must
consider the possibility of systematic uncertainties that could affect
our results.  We investigate the accuracy of the ASAS-SN photometry
and derived brightness trends with various comparison samples.

\subsubsection{Stars in the Same Field as Boyajian's Star}
\label{samefield}

\begin{figure*}[th!]
\epsscale{1.20}
\plotone{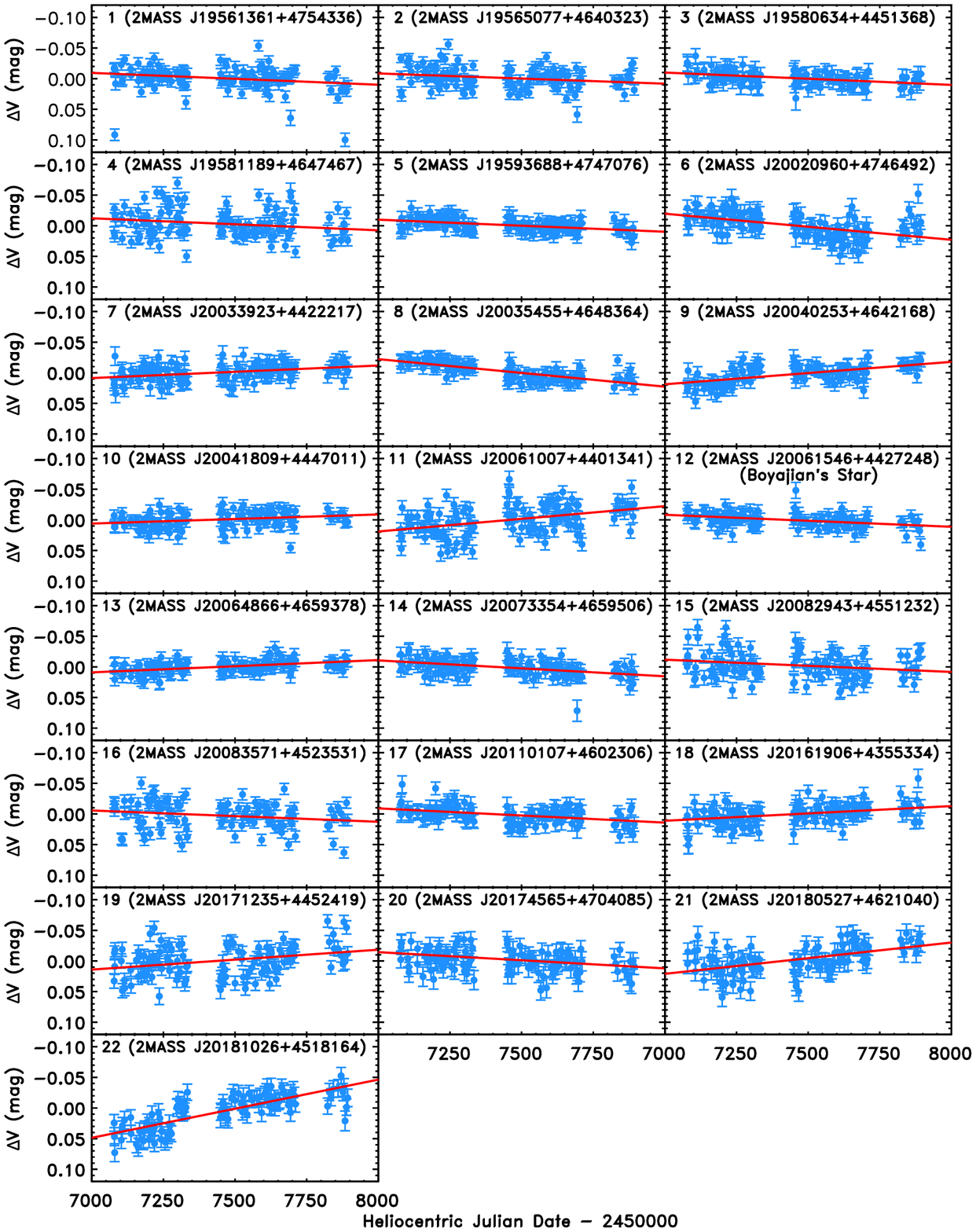}
\caption{ASAS-SN light curves of the 22 stars (out of 1124 total with
  $11.5 < V < 12.5$) in the Boyajian's Star ASAS-SN field observed
  with the bd camera for which we detect brightness trends at
  $>4\sigma$.  The median magnitude of each star has been subtracted
  from all measurements for display purposes.  Boyajian's Star
  (star 12, 2MASS~J20061546+4427248) is the fourth panel in the right-hand
  column.}
\label{trends}
\end{figure*}

Our first comparison sample consists of 1124 stars in the SIMBAD
database \citep{simbad} within $\sim0.5$~mag of Boyajian's Star ($11.5
< V < 12.5$) that fall in the same ASAS-SN field and camera as
Boyajian's Star.  There are at least 30 ASAS-SN measurements passing
our quality cuts (see \S\ref{tabbysstar}) for 1078 of these stars.  We
carry out the same OLS fit described in \S\ref{tabbysstar} for these
stars, finding a total of 21 (in addition to Boyajian's Star) that
have best-fit linear trends detected at greater than $4\sigma$
significance and a scatter around the trend of less than 0.03~mag.
The latter cut removes stars with large amplitude variability, of
which there are 45 examples identified by visual examination of the
ASAS-SN light curves and Lomb-Scargle periodograms of the brightness
measurements.  These 21 stars (and Boyajian's Star) are listed in
Table~\ref{trendstars_table} and their light curves are displayed in
Figure~\ref{trends}.  Unlike for Boyajian's Star, a linear trend is
clearly not a sufficient description for the behavior of several of
these stars (e.g., stars 6, 9, and 22 in
  Table~\ref{trendstars_table}), but a simple linear fit to the
available ASAS-SN photometry should still provide a reasonable
criterion for selecting variable stars.  Finding 21 stars with
$>4\sigma$ trends significantly exceeds the number expected from
random errors if the ASAS-SN photometric noise is Gaussian.  These
stars could either be previously unknown variables or be pointing to
low-level systematic uncertainties in the data.  Only one of the stars
(star 1, which is an Algol-type eclipsing binary;
\citealt{devor08}) is listed as a known variable in the AAVSO
\citep{watson06} or SIMBAD \citep{simbad} databases.  However, the
lack of existing variable identifications should not be taken as
evidence against variability; many public wide-field surveys either
avoid regions of the sky so close to the Galactic plane (e.g., SDSS;
\citealt{york00}, Catalina Real-Time Transient Search;
\citealt{drake09}) or are saturated for such bright stars (e.g.,
Pan-STARRS; \citealt{magnier13}).

\input{trendstar_table_4sigma.tex}

Unfortunately, because Boyajian's Star is located in the corner of the
\emph{Kepler} field of view there is very limited overlap between the
ASAS-SN field containing the star and the \emph{Kepler} field.  Just
five of the 21 candidate low-amplitude variables from ASAS-SN
(stars 2, 3, 4, 7, and 10) have \emph{Kepler} long
cadence light curves, and only star 7 shows
significant short-term variability.  We use the methodology described
by \citet{montet17} to search for longer-period variability of these
stars in the \emph{Kepler} full frame images (FFIs).  All five stars
display clear long-term trends (of smaller amplitude than seen in the
ASAS-SN data), which we regard as possible stellar activity signals.

\begin{figure}[th!]
\epsscale{1.16}
\plotone{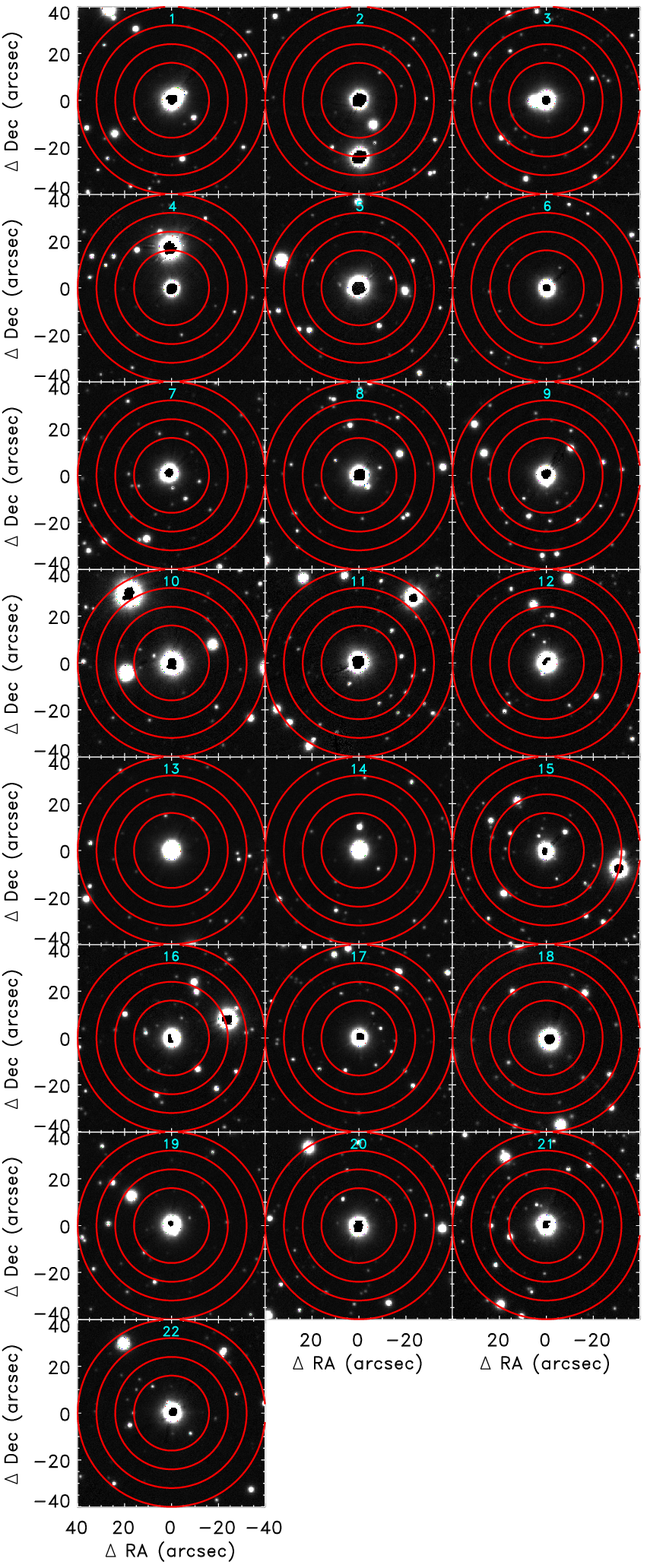}
\caption{Pan-STARRS1 r-band images \citep{chambers17,ps1_images} of
  all stars in the Boyajian's Star field with brightness trends
  detected at $>4\sigma$.  Each image is $80\arcsec \times 80\arcsec$,
  and the 2, 3, 4, and 5 ASAS-SN pixel photometric apertures are
  displayed as red circles.  Note that the cores of most of these
  stars are saturated in the Pan-STARRS data.  Boyajian's Star
  (star 12) is shown in the fourth panel in the right-hand
  column.  }
\label{images}
\end{figure}

We also evaluate the variability of this sample of candidate variables
by considering their behavior as a function of the aperture radius
used for ASAS-SN photometry.  We compare the results for our standard
3-pixel radius aperture with apertures of 2, 4, and 5 pixels (see
Figure~\ref{images}).  For seven of the stars the ASAS-SN trend is
only statistically significant for one or two choices of aperture.  We
conclude that blending with other stars located near the boundary of
the photometric aperture is likely responsible for the apparent
variability of these stars.\footnote{For stars 2, 4, 10, 15, and 16
  the nearby bright stars causing variable blending
  in the different apertures are obvious in Fig.~\ref{images}.  For
  stars 1 and 18 the origin of
  the differing behavior as a function of aperture size is less
  evident, although star 1 has a close
  ($\sim2\arcsec$) blended companion visible to the northwest.
  Star 11 also has a similarly bright neighbor that
  affects the outermost aperture, but its variability is quite clear
  in the smaller apertures.}  We remove them from further
consideration, leaving 14 possible variables.  Five of the remaining
stars show substantially decreasing significance as the aperture
radius increases, which again indicates blending is occurring but is
still consistent with the star at the center of the aperture being
variable.

A powerful new resource for detecting photometric variability across
the entire sky is Gaia \citep[e.g.,][]{eyer17,belokurov17,deason17}.
While variability information is not included in the first Gaia data
release \citep{gaiadr1}, \citet{belokurov17} showed that it is
possible to identify variable stars by comparing the reported flux
uncertainties in the Gaia DR1 catalog to the expected photon noise.
We calculate the Amplitude metric defined by \citet{belokurov17} for
the fourteen candidate variables in the same ASAS-SN field as
Boyajian's Star and find that eleven easily satisfy the conservative
variability threshold of $\rm{Amp} > 0.22G - 4.87$ set by
\citeauthor{belokurov17} This result suggests that at least
$\sim70$\%\ of the stars with $>4\sigma$ trends in ASAS-SN are
actually variable.  Two additional stars in this sample have
photometric amplitudes above but very close to the threshold.
Interestingly, Boyajian's Star itself is not clearly variable in the
Gaia data, with an amplitude slightly below the \citet{belokurov17}
threshold.  Gaia DR1 includes observations from 2014 July to 2015
September \citep{gaiadr1}, and it is possible that the star was
relatively quiescent over this period.

\subsubsection{Stars in the \emph{Kepler} Field}
\label{kepfield}

In order to provide a larger sample of stars observed by both ASAS-SN
and \emph{Kepler} we use the bd camera data for a neighboring ASAS-SN
field that falls entirely within the boundary of the \emph{Kepler}
field.  There are 1166 SIMBAD stars in this field in the same
magnitude range used above.  Repeating the light curve fitting
exercise for these stars yields 26 with $>4\sigma$ brightness trends
out of 944 with at least 30 good ASAS-SN measurements.  Although the
ASAS-SN observations of this field cover $1000-1200$~days instead of
the $\sim800$~days for the field containing Boyajian's Star, we limit
the light curve fits to the same time span for consistency.  Thirteen
of the stars with ASAS-SN trends have \emph{Kepler} data (the others
fall in gaps between \emph{Kepler} CCDs).  We detect variability in
the \emph{Kepler} FFI data for 8 of the 13 remaining stars.  In
combination with the five stars with FFI measurements in the
Boyajian's Star field (\S\ref{samefield}), \emph{Kepler} observations
indicate that at least $\sim70$\%\ of stars with $>4\sigma$ ASAS-SN
trends are astrophysically variable.  Many of the stars are strongly
spotted, showing $>1\%$ variability from spot modulation, and a number
exhibit long-term trends that may be due to magnetic cycles.  The
character of the variability does not necessarily match between the
two data sets, at least in part because of the limited temporal
overlap.  Nevertheless, we consider this result to be confirmation
that most stars with $>4\sigma$ brightness trends in the ASAS-SN
photometry are genuinely variable, regardless of the nature of that
variability.

\subsection{Significance of ASAS-SN Variability}

If we were to assume that \emph{all} of the ASAS-SN trend stars except
for Boyajian's Star are false positives --- i.e., that none of them
are actually variable --- then the false positive rate would be
1.9\%\ in the Boyajian's Star field and the detection of long-term
variability in Boyajian's Star would have a confidence level of
98.1\%.  Given the results above from comparisons with Gaia and
\emph{Kepler} (\S\S\ref{samefield} and \ref{kepfield}) we conclude
that the actual false positive rate is $\le 0.6\%$.  We therefore
estimate a confidence level for the ASAS-SN detection of variability
in Boyajian's Star of $\ge 99.4\%$.  This confidence level corresponds
to $\sim3\sigma$ significance, lower than determined using the
statistical uncertainties in the data alone, but still providing
strong support for continued long-term brightness changes of the star.

\section{ANALYSIS OF ASAS DATA}
\label{asas}

\begin{figure*}[th!]
\epsscale{1.185}
\plotone{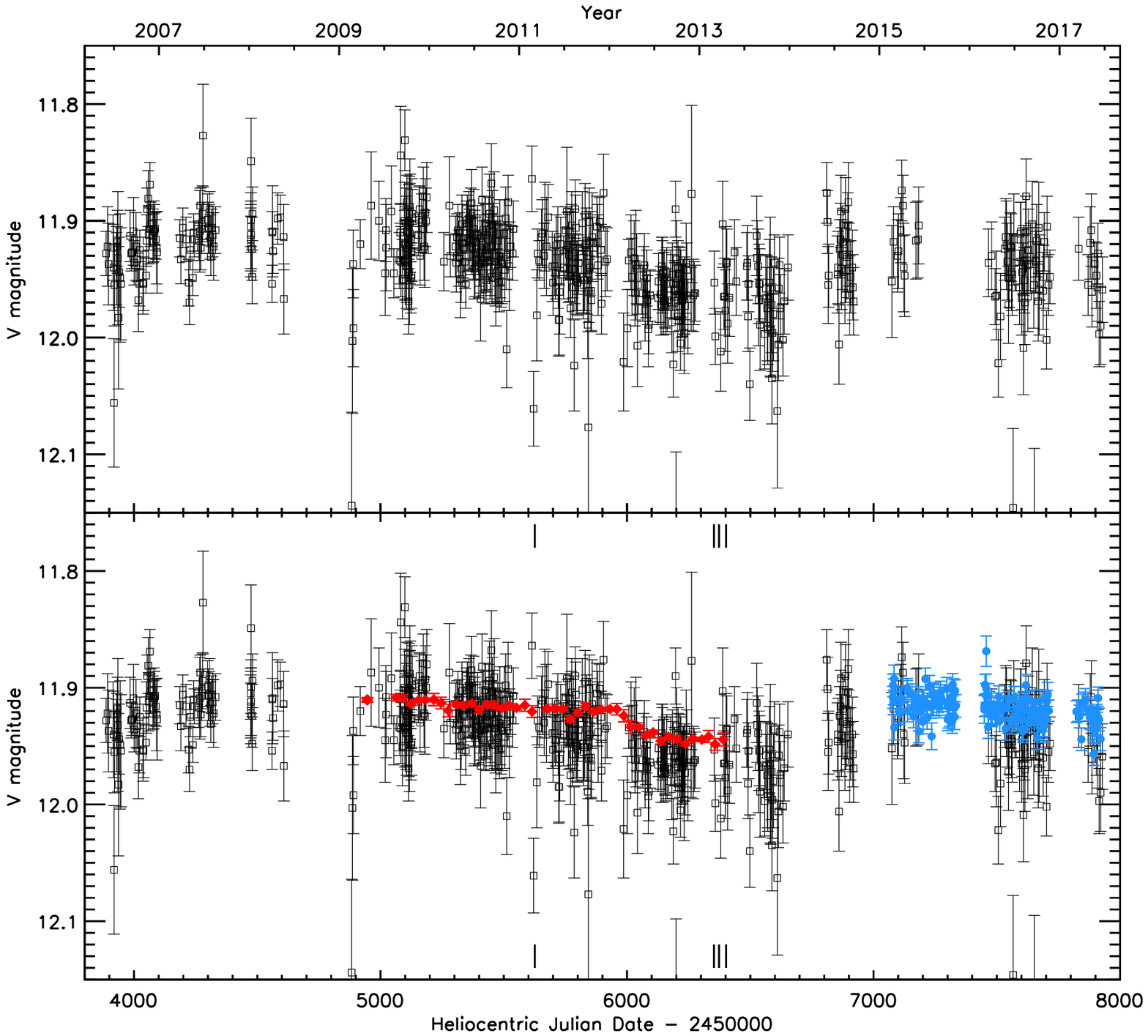}
\caption{(top) ASAS light curve of Boyajian's Star from 2006 to
  present.  (bottom) ASAS light curve with \emph{Kepler} FFI light
  curve (red diamonds) from \citet{ms16} and ASAS-SN light curve (blue
  circles) overlaid.  The FFI light curve is uncalibrated, so we
  applied a zero-point offset of 11.92 mag in order to match the ASAS
  magnitude at the beginning of the \emph{Kepler} mission.  For
  display purposes we also offset the ASAS-SN magnitudes by
  $+0.01$~mag to bring them into agreement with the ASAS magnitude in
  early 2015 ($\rm{HJD} \sim 2457100$).  The black tick marks near the
  top and bottom of the lower panel indicate the times of the
  strongest dips ($>2\%$ flux decrement) seen by \emph{Kepler}.}
\label{asas_lc}
\end{figure*}

The ASAS $V$-band light curve of Boyajian's Star is displayed in
Figure~\ref{asas_lc}, and the measurements are listed in
Table~\ref{asas_data_table}.  The appearance of the light curve is
dominated by a dimming of $\sim0.04$~mag from $\rm{JD} = 2455000$ to
$\rm{JD} = 2456500$.  We show in the bottom panel of
Fig.~\ref{asas_lc} that this decrease in brightness aligns perfectly
with the fading detected in the \emph{Kepler} FFIs by \citet{ms16}.
Strikingly, though, the star does not continue fading from there, but
rather returns to its previous brightness level by $\rm{JD} \approx
2457000$.  Unfortunately, there are few ASAS measurements available in
the intervening 1.5~yr to reveal exactly when and how this brightening
took place, but its duration is constrained to be $\lesssim200$~d.
From $\rm{JD} \approx 2457100$ (early 2015) to the present we measure
a dimming rate of $15 \pm 6$~mmag~yr$^{-1}$, consistent with the
dimming seen in the ASAS-SN observations at the $\sim1.4\sigma$ level.
The significance of the ASAS fading over this time interval is lower
than in ASAS-SN because of the larger photometric uncertainties and
the slower observational cadence.

\input{asas_data_table_stub.tex}

At the beginning of the ASAS time series there is some evidence of
another brightening event, with the star's magnitude changing from $V
= 11.93$ at $\rm{JD} < 2454000$ to $V = 11.91$ by $\rm{JD} = 2454500$.
In this case the brightening appears to extend over $\sim200$~d.
Thus, rather than monotonically fading as has been inferred from the
photometry of \citet{schaefer16} and \citet{ms16}, the star has likely
both brightened and dimmed measurably over the last decade.  This
result does not necessarily contradict the century-long dimming seen
by \citet{schaefer16}, which would only amount to 0.018~mag over the
ASAS baseline.  Variability at the level of a few percent on
timescales of a few years such as we measure in ASAS-SN and ASAS would
not be detectable in the DASCH photometry \citep{grindlay09,laycock10}
used by \citet{schaefer16} given the much larger uncertainties in the
photographic data.

If we fit the entire ASAS light curve with a simple linear model as in
\S\ref{tabbysstar}, we derive an overall dimming rate of $3.6 \pm
0.4$~mmag~yr$^{-1}$.  This fading rate is nearly identical to that
measured by \citet{ms16} over the first $\sim1000$~d of the
\emph{Kepler} mission.  As discussed above, it is apparent from
Fig.~\ref{asas_lc} that a simple linear model is not a good
description of the data for this longer time span.  However, the fact
that the slope of the fit is significantly different from zero
provides further evidence for the long-term photometric variability of
Boyajian's Star.

The long-term dimming rate in ASAS over the duration of the
\emph{Kepler} observations is $15.8 \pm 1.5$~mmag~yr$^{-1}$, almost a
factor of two larger than in the \citet{ms16} \emph{Kepler}
measurements.  Because the \emph{Kepler} bandpass extends to
significantly longer wavelengths than a standard $V$ filter, this
difference could be a sign that the dimming is weaker in the red (also
see the Appendix).  Either dust or a change in stellar temperature
could cause the star to get redder as it dims.

We also analyzed the same sample of 1124 stars of similar brightness
described in \S\ref{samefield} using ASAS data.  Of these stars, 1117
have at least 30 good ASAS measurements, and we find 85 stars with
best-fit linear trends detected at greater than $4\sigma$ significance
with a scatter around the trend of less than 0.05~mag (to allow for
the larger ASAS uncertainties for each measurement).  Boyajian's Star
has the tenth most significant trend and the eleventh largest slope
among this sample.  Stars 2, 3, 5, 6, 7, 19, 21, and 22 from 
Table~\ref{trends} also
exhibit $>4\sigma$ trends in both the ASAS-SN and ASAS data sets,
supporting the identification of these stars as variable.

Given that the ASAS data set covers the full time span of the
\emph{Kepler} light curve of Boyajian's Star, it is possible that
there could be ground-based detections of some of the dips discovered
by \citet{boyajian16}.  Figure~\ref{asas_lc} shows that there are more
ASAS outliers to fainter magnitudes than brighter ones, consistent
with the possibility of dipping activity.  However, none of those data
points match the times of the dips seen with \emph{Kepler}.  There is
a near coincidence between the deep (16\%) D800 dip and an ASAS
measurement at $\mathrm{HJD} = 2455621.16839$ that is more than
$2\sigma$ below the mean magnitude, but the two are separated by more
than 4 days and the \emph{Kepler} dip was very short, so this ASAS
measurement must not be accurate.  Nevertheless, there may be genuine
dips in the ASAS data at other times; most notably at $\mathrm{HJD}
\approx 2454900$ (immediately before \emph{Kepler} observations began)
and in mid-2016 ($\mathrm{HJD} \approx 2457600$).

\begin{figure}[t!]
\epsscale{1.24}
\plotone{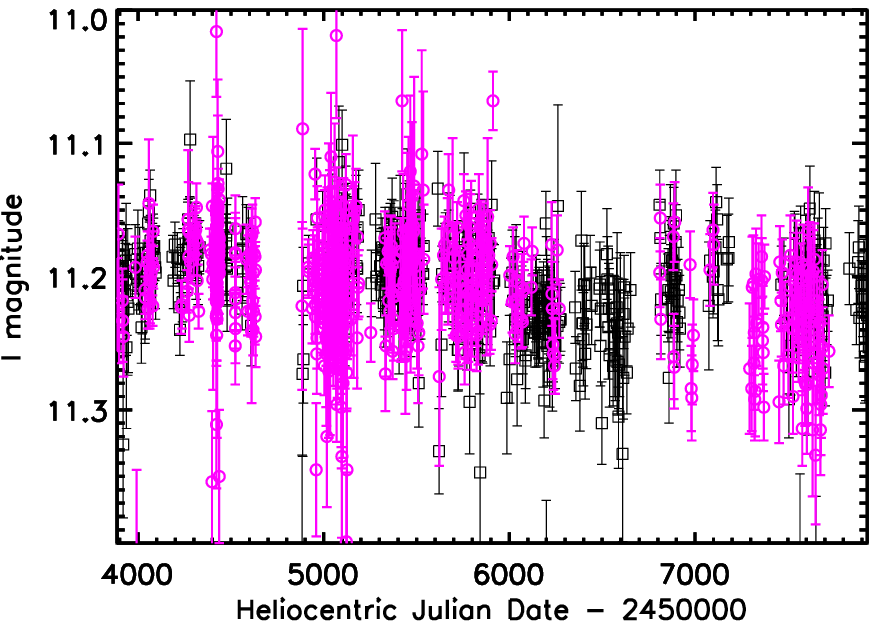}
\caption{(top) ASAS light curve of Boyajian's Star, spanning from 2006
  to present.  The $V$-band data from Fig.~\ref{asas_lc}, offset by
  $-0.73$~mag to match the median $I$-band magnitude, are displayed as
  open black squares and the $I$-band data are plotted as magenta
  circles.}
\label{asas_i}
\end{figure}

Boyajian's Star is detected as variable in the ASAS $I$-band
photometry as well, with a dimming rate of $4.6 \pm
0.4$~mmag~yr$^{-1}$ from 2006 June to 2016 November.  The sequential
brightening and fading that is evident in the $V$-band light curve is
not obvious in $I$; in particular, there is no sign of the brightening
between the $2009-2013$ \emph{Kepler} observations and the
$2015-\rm{present}$ ASAS-SN measurements that is seen in $V$ (see
Figure~\ref{asas_i}).

\section{DISCUSSION}
\label{discussion}

The results of \citet{schaefer16} and \citet{ms16} have generally been
assumed to indicate that the brightness of Boyajian's Star is
monotonically decreasing with time.  While the ASAS-SN measurements
are consistent with that picture, the longer time baseline afforded by
the ASAS data demonstrates that in fact the star's variability may be
more complicated, with multiple slow dimming and brightening episodes
over the past decade.  This finding may have significant implications
for models to explain the behavior of the star.

Published ideas to account for the unusual changes in the brightness
of Boyajian's Star generally suffer from one of two flaws: they are
either statistically unlikely \citep[e.g.,][]{metzger17} or they
explain only one aspect of the observed flux variations
\citep[e.g.,][]{bq16,mg16,sheikh16,nb17,ballesteros17,katz17}.  The
hypotheses that can potentially account for both the brief dips and
the long-term variability include the consumption of a planet
\citep{metzger17}, an internal obstruction of heat flux in the star
\citep{foukal17}, or an intervening disk-bearing object \citep{ws16}.
In the first two cases, the proposed explanation relies on the star's
luminosity monotonically decreasing.  If Boyajian's Star also
brightens at times then these models as presented may not be viable.
For the planetary merger hypothesis, minor brightening events
interrupting the long-term dimming could be the result of the star
ingesting additional bodies.  The collision of a moon-sized object
with the star would cause brightening of approximately the correct
magnitude (K. Shen \& B. Metzger 2017, personal communication).
However, the \citet{metzger17} models indicate that the brightening
from such an event should occur very quickly, which appears to be in
conflict with the observed duration of the $2006-2007$ brightening
(\S~\ref{asas}).

The most obvious culprit for long-term dimming and brightening of
Boyajian's Star is a magnetic activity cycle.  The two brightening
episodes seen in the ASAS photometry occur around $\rm{JD} = 2454000$
and $\rm{JD} = 2456800$, suggesting an $\sim8$~yr cycle length.  If
this brightening is cyclical (whether from stellar activity or
otherwise) then the star should brighten again near $\rm{JD} =
2459600$ (2022 January), with an uncertainty of a few months.

The interpretation of the photometric variability as a stellar cycle
must confront the question of whether the amplitude, time scale, and
even presence of magnetic activity in such a star has any precedent.
In the compilation of stars tracked in Ca H\&K activity indicators,
\citet{Egeland17} lists 28 stars with colors $0.5 < B-V < 0.6$; of
these, only 9 show cycles, ranging from ``good" to ``poor" quality.
The mean (median) cycle period in this hot-star sample is $\sim
11.3$~(10.0) years. Of the 19 stars without cycles, only 5 of them are
``flat", the rest showing variability or long-term trends.
\citet{Mathur14} noted a possible magnetic cycle with
$\textrm{P}>1400$ days and the signature of differentially rotating
spots in the seismically well-studied KIC~3733735, which is also a hot
($\sim 6700$~K) rapidly rotating ($\textrm{P} = 2.5$ days) early
F-type star. Boyajian's Star shows similar photometric evolution over
the course of the \emph{Kepler} mission (see, e.g., Fig.~3 of
\citealt{boyajian16}).  Moreover, the increase in short-term
photometric variability occurs simultaneously with the observed
decrease in the star's brightness, similar to the correlation observed
in \emph{Kepler} observations of rapidly rotating sun-like stars
\citep{montet17}.\footnote{This simultaneity presumes that the 0.88~d
  rotation signature is from KIC~8462852, which the astrometric shifts
  detected by \cite{mg16} suggest is not the case.}  Although
Boyajian's Star is close to the boundary where we expect deep surface
convection zones (and thus magnetic activity) to disappear, stellar
models for stars of $6500-6700$~K at solar metallicity still have
convection zones of order a few to 10\% of their radii
\citep{vanSaders13} and Rossby numbers \citep[as calculated
  in][]{vanSaders16} generally less than 2.0 assuming a 0.88~d
rotation period.  It is unknown exactly when dynamo action in such
convection zones would cease \citep[but see][]{Metcalfe17}.

Although the presence of magnetic activity and observed cycle lengths
do have some precedent, the photometric amplitude of the cycle is more
puzzling. The luminosity variations are much larger than seen for
other similar stars \citep[e.g.,][]{ms16}. In the \citet{Lockwood07}
compilation of stars with both Ca H\&K and photometric monitoring,
stars with $B-V < 0.6$ have long-term variability that is
typically less than $\sim 0.01$ mag in the Str\"{o}mgren b and y
filters.  Moreover, magnetic activity offers no explanation for the
brief dips Boyajian's Star undergoes.

Another plausible, although perhaps less natural, explanation for the
ASAS variability is changing line-of-sight absorption by the
interstellar medium \citep{ws16}.  The overall reddening inferred for
the star is $\rm{E(}B-V\rm{)} = 0.11 \pm 0.03$~mag \citep{boyajian16}, so
variations in the V-band extinction of $\sim0.03$~mag are not
impossible.  However, while measurable changes in the ISM absorbing
column on timescales of years are known
\citep[e.g.,][]{hobbs82,price00,wf01,smith13,galazutdinov13},
detectable flux decrements from varying ISM absorption are not, and
lines of sight with the necessary small-scale structure to cause such
changes are rare.  This idea certainly cannot be ruled out at present,
but the more complicated the photometric behavior of Boyajian's Star
becomes, the more contrived the ISM structure must be in order to
explain it.  Requiring large ISM density contrasts on spatial scales
sufficiently small to account for the dips may run afoul of the
argument by \citet{lacki16} that the cause of the photometric
anomalies must be surprisingly common in order for \emph{Kepler} to
have been likely to detect such a star.  High-resolution monitoring of
the Na~D and \ion{K}{1} absorption lines in the spectrum of Boyajian's
Star over several years can provide a strong test of this hypothesis.

\section{SUMMARY AND CONCLUSIONS}
\label{conclusions}

We have used long-term survey observations of the \emph{Kepler} field
by ASAS-SN and ASAS to investigate the brightness of Boyajian's Star
from 2006 to 2017.  Over the past two years, we detect dimming in both
surveys.  With the higher-precision ASAS-SN photometry we measure a
decline rate of $6.3 \pm 1.4$~mmag~yr$^{-1}$.  Fading nearly identical
to what we measure has been detected independently using observations
from the Hereford Arizona Observatory (B. Gary 2017, personal
communication), and similar results have also just been reported by
\citet{meng17}.  Prior to the beginning of ASAS-SN coverage of this
location, ASAS shows that the star brightened in 2014, faded as
previously seen with \emph{Kepler} from 2009-2013, and may have
brightened in 2006.  We therefore conclude that the variability of
Boyajian's Star is likely not monotonic as has been assumed until now.

The existence of periods in which Boyajian's Star brightens noticeably
and the possibility of cyclic variability significantly change the
long-term photometric signature that models of the star's behavior
have sought to explain.  Most published models aim to account for the
brief dips discovered by \citet{boyajian16} without addressing the
long-term dimming measured by \citet{schaefer16} and \citet{ms16}.
The \citet{metzger17} hypothesis that the star has recently consumed a
planet predicts a strictly declining luminosity after $\sim1$~yr past
the merger, which does not appear to be consistent with the ASAS
photometry, although more complicated versions of this idea may not be
ruled out.  We examine whether the detected brightness variations
could be the result of stellar activity, but find that explanation to
be unsatisfying because of the unusually large amplitude of the
variability and the difficulty of accounting for the brief dips with
the same mechanism.  Interstellar absorption could conceivably be
responsible for both the dips and long-term flux variations, but only
with unprecedented levels of small-scale ISM structure.

Until the recent study by \citet{meng17}, all detections of both the
long-term dimming of Boyajian's Star and the brief dips in brightness
have relied on photometry in a single band \citep{schaefer16,ms16}.
Although we find that the ASAS $I$-band observations indicate fading
as well, the larger uncertainties of those measurements prevent us
from confidently detecting any color changes in the star over the past
decade.  The key question of how \emph{all} of the brightness changes
observed in Boyajian's Star vary with wavelength therefore remains
unanswered.  We strongly encourage high-precision multi-year
monitoring efforts in multiple photometric bands that will be
sensitive enough to detect changes in the star's brightness of
$1-2$\%.  Multi-color imaging obtained over the time interval when
Boyajian's Star will be observed by the TESS mission \citep{tess}
would be particularly valuable.  The Zwicky Transient Facility
\citep{bk17} and the Gaia satellite \citep{gaia16} offer two possible
avenues for pursuing a long-term, well-calibrated monitoring program
covering a range of wavelengths, but the results presented here
demonstrate that smaller-scale surveys are capable of achieving the
required accuracy as well.

\acknowledgements{We thank Jason Wright, Ken Shen, Brian Metzger,
  Tabby Boyajian, Bruce Gary, Ryan Oelkers, Rob Siverd, Johanna Teske,
  Eddie Schlafly, and Jim Davenport for helpful conversations, and
  Huan Meng and collaborators for sharing a draft of their paper prior
  to publication.

This research has made use of NASA's Astrophysics Data System
Bibliographic Services and the SIMBAD database, operated at CDS,
Strasbourg, France.  This research also made use of the International
Variable Star Index (VSX) database, operated at AAVSO, Cambridge,
Massachusetts, USA.

We thank the Las Cumbres Observatory and its staff for its continuing
support of the ASAS-SN project. ASAS-SN is supported by the Gordon and
Betty Moore Foundation through grant GBMF5490 to the Ohio State
University and NSF grant AST-1515927.  Development of ASAS-SN has been
supported by NSF grant AST-0908816, the Mt. Cuba Astronomical
Foundation, the Center for Cosmology and AstroParticle Physics at the
Ohio State University, the Chinese Academy of Sciences South America
Center for Astronomy (CASSACA), the Villum Foundation and George
Skestos.

BJS is supported by NASA through Hubble Fellowship grant
HST-HF-51348.001 awarded by the Space Telescope Science Institute,
which is operated by the Association of Universities for Research in
Astronomy, Inc., for NASA, under contract NAS 5-26555.  Work by BTM
was performed under contract with the California Institute of
Technology/Jet Propulsion Laboratory funded by NASA through the Sagan
Fellowship Program executed by the NASA Exoplanet Science Institute.
CSK is supported by NSF grants AST-1515927 and AST-1515876.  TW-SH is
supported by the DOE Computational Science Graduate Fellowship, grant
number DE-FG02-97ER25308.

This research was made possible through the use of the AAVSO
Photometric All-Sky Survey (APASS), funded by the Robert Martin Ayers
Sciences Fund.  APASS is also supported by NSF grant AST-1412587.
This research has made use of data provided by Astrometry.net
(\citealt{lang10}).

The Pan-STARRS1 Surveys (PS1) and the PS1 public science archive have
been made possible through contributions by the Institute for
Astronomy, the University of Hawaii, the Pan-STARRS Project Office,
the Max-Planck Society and its participating institutes, the Max
Planck Institute for Astronomy, Heidelberg and the Max Planck
Institute for Extraterrestrial Physics, Garching, The Johns Hopkins
University, Durham University, the University of Edinburgh, the
Queen's University Belfast, the Harvard-Smithsonian Center for
Astrophysics, the Las Cumbres Observatory Global Telescope Network
Incorporated, the National Central University of Taiwan, the Space
Telescope Science Institute, the National Aeronautics and Space
Administration under Grant No. NNX08AR22G issued through the Planetary
Science Division of the NASA Science Mission Directorate, the National
Science Foundation Grant No. AST-1238877, the University of Maryland,
Eotvos Lorand University (ELTE), the Los Alamos National Laboratory,
and the Gordon and Betty Moore Foundation.  }

{\it Facilities:} ASAS-SN, ASAS 

\appendix
\section{EXISTING MULTICOLOR PHOTOMETRY OF BOYAJIAN'S STAR}
\label{color}

The color dependence of the variability of Boyajian's Star would
provide a critical clue to the origin of the dimming on both short and
long timescales.  If light from the star is being obscured by dust, it
would get redder as it dims, while if the light is being blocked by
macroscopic objects such as rocky collisional debris
\citep[e.g.,][]{boyajian16} or artificial structures \citep{wright16}
then the dimming would be independent of wavelength.  In order to
explore whether any changes in the color of the star can be detected
in existing data, we collect all available optical photometry from the
literature and public catalogs in Table~\ref{photom_table}.
Unfortunately, there are significant disagreements between different
references, especially in $B$ and $V$, where the total range of the
reported magnitudes is more than 0.25~mag.  The $B - V$
colors, though, are generally consistent.  The largest outliers in
these bands are the Kepler INT Survey \citep{greiss12a,greiss12b} and
\citet{boyajian16}.  Since the magnitude discrepancies are much larger
than the amplitude of the $V$-band variability presented in
\S\S~\ref{analysis} and \ref{asas}, we suspect that they result from
inconsistent calibrations or magnitude systems rather than
astrophysical changes.

\input{photom_table.tex}

Of the first five data sets listed in Table~\ref{photom_table}, only
APASS offers any time resolution, with four measurements in each band
during 2010 June and a single observation in 2012 October, spanning
the rapid decrease in the brightness of Boyajian's Star detected by
\citet{ms16}.  Averaging the 2010 measurements together, we find
dimming of 0.35~mag in $B$, 0.11~mag in $V$, 0.13~mag in $g$, 0.06~mag
in $r$, and 0.16~mag in $i$ from 2010 to 2012.  Based on the overall
uncertainties for each band listed in Table~\ref{photom_table}, we
assume that each APASS magnitude has an uncertainty of $\sim0.05$~mag,
so the uncertainties on the magnitude differences should be
$\sim0.07$~mag.  This $V$ and $g$ dimming is larger than what we
measure with ASAS by $\sim1\sigma$.  Although the decrease in
brightness in $B$ seems anomalously large relative to the other
filters for $\rm{R}_V = 3.1$ dust, the APASS data provide some
evidence that Boyajian's Star reddened during the course of the rapid
dimming event in early 2012.

\bibliographystyle{apj}

\end{document}

%% file: asassn_data_table_stub.tex
\begin{deluxetable}{lcccc}
\tablecaption{ASAS-SN Photometry of Boyajian's Star}
\tablewidth{0pt}
\tablehead{
\colhead{HJD} &
\colhead{$V$} &
\colhead{$\Delta V$} & 
\colhead{FWHM} &
\colhead{Zero point\tablenotemark{a}} \\
\colhead{} &
\colhead{[mag]} &
\colhead{[mag]} &
\colhead{[pix]} &
\colhead{[mag]}
}
\startdata
2457078.15792 & 11.898 & 0.009 & 1.56 & -2.26 \\
2457079.15672 & 11.888 & 0.010 & 1.65 & -2.27 \\
2457080.15999 & 11.903 & 0.010 & 1.50 & -2.24 \\
2457081.16257 & 11.894 & 0.011 & 1.69 & -2.25 \\
2457083.16491 & 11.882 & 0.012 & 1.73 & -2.25 \\
2457084.16041 & 11.924 & 0.011 & 2.12 & -2.26 \\
2457103.14733 & 11.910 & 0.010 & 1.51 & -2.20 \\
2457106.11872 & 11.912 & 0.010 & 1.54 & -2.22 \\
2457107.13016 & 11.899 & 0.009 & 1.47 & -2.26 \\
2457114.09387 & 11.902 & 0.010 & 1.66 & -2.26 
\enddata
\tablecomments{This table is available in its entirety in the electronic
edition of the journal.  A portion is reproduced here to provide
guidance on form and content.}
\tablenotetext{a}{The ASAS-SN zero point (ZP) of each image
is defined such that a source with a flux of one count has a
magnitude of $25 + \textrm{ZP}$.}
\label{asassn_data_table}
\end{deluxetable}

%% file: trendstar_table_4sigma.tex
\begin{deluxetable*}{llcccccc}
\tablecaption{ASAS-SN Stars With Significant Trends}
\tablewidth{0pt}
\tablehead{
\colhead{Star} &
\colhead{ID} &
\colhead{$V$} &
\colhead{$B-V$} &
\colhead{$V-K$} &
\colhead{Brightness Change} &
\colhead{Scatter} & 
\colhead{$\chi_{{\rm red}}^{2}$} \\
\colhead{} &
\colhead{} &
\colhead{} &
\colhead{mag} &
\colhead{mag} &
\colhead{mmag yr$^{-1}$} &
\colhead{mag} &
\colhead{}
}
\startdata
1 & 2MASS J19561361+4754336 & 11.24 & 0.32 & 0.41 & \phn\phs$  7.0 \pm 1.2$ & 0.020 & 4.63 \\
2 & 2MASS J19565077+4640323 & 11.89 & 1.28 & 2.97 & \phn\phs$  5.8 \pm 1.2$ & 0.017 & 4.17 \\
3 & 2MASS J19580634+4451368 & 11.91 & 0.79 & 1.68 & \phn\phs$  7.9 \pm 1.9$ & 0.010 & 0.58 \\
4 & 2MASS J19581189+4647467 & 12.26 & 0.42 & 1.61 & \phn\phs$  7.0 \pm 1.4$ & 0.023 & 5.71 \\
5 & 2MASS J19593688+4747076 & 11.55 & 1.72 & 3.82 & \phn\phs$  6.9 \pm 1.5$ & 0.009 & 0.65 \\
6 & 2MASS J20020960+4746492 & 12.19 & 0.83 & 1.72 &     \phs$ 16.2 \pm 1.8$ & 0.017 & 1.69 \\
7 & 2MASS J20033923+4422217 & 12.15 & 0.55 & 1.39 & \phn$   -  7.7 \pm 1.6$ & 0.013 & 1.18 \\
8 & 2MASS J20035455+4648364 & 11.75 & 0.84 & 1.97 &     \phs$ 16.5 \pm 1.5$ & 0.009 & 0.93 \\
9 & 2MASS J20040253+4642168 & 11.93 & 0.75 & 1.74 &     $   - 13.2 \pm 1.5$ & 0.013 & 1.46 \\
10 & 2MASS J20041809+4447011 & 11.42 & 0.63 & 1.59 & \phn$   -  5.8 \pm 1.3$ & 0.013 & 2.02 \\
11 & 2MASS J20061007+4401341 & 11.77 & 0.98 & 2.16 &     $   - 14.9 \pm 1.5$ & 0.022 & 4.08 \\
12 & 2MASS J20061546+4427248 & 11.85 & 0.51 & 1.35 & \phn\phs$  6.3 \pm 1.4$ & 0.011 & 1.08 \\
13 & 2MASS J20064866+4659378 & 11.82 & 0.63 & 1.54 & \phn$   -  7.3 \pm 1.4$ & 0.009 & 0.91 \\
14 & 2MASS J20073354+4659506 & 12.28 & 0.75 & 1.79 & \phn\phs$  9.2 \pm 1.6$ & 0.011 & 0.95 \\
15 & 2MASS J20082943+4551232 & 12.47 & 0.61 & 1.32 & \phn\phs$  8.3 \pm 1.7$ & 0.021 & 3.31 \\
16 & 2MASS J20083571+4523531 & 12.15 & 0.54 & 1.05 & \phn\phs$  6.7 \pm 1.3$ & 0.022 & 5.61 \\
17 & 2MASS J20110107+4602306 & 12.25 & 0.80 & 1.71 & \phn\phs$  8.4 \pm 1.6$ & 0.013 & 1.25 \\
18 & 2MASS J20161906+4355334 & 12.27 & 1.35 & 3.29 & \phn$   -  8.7 \pm 1.7$ & 0.015 & 1.54 \\
19 & 2MASS J20171235+4452419 & 12.12 & 0.68 & 1.86 &     $   - 11.9 \pm 1.6$ & 0.024 & 4.54 \\
20 & 2MASS J20174565+4704085 & 11.83 & 0.69 & 1.94 & \phn\phs$  9.7 \pm 2.1$ & 0.017 & 1.44 \\
21 & 2MASS J20180527+4621040 & 12.00 & 0.48 & 1.53 &     $   - 18.5 \pm 2.4$ & 0.020 & 1.74 \\
22 & 2MASS J20181026+4518164 & 12.20 & 0.62 & 1.57 &     $   - 35.2 \pm 2.1$ & 0.018 & 1.60 
\enddata
\tablecomments{$B$ and $V$ magnitudes are taken from APASS \citep{Henden14},
  with the exception of star 4, which is not included in the APASS DR9
catalog (presumably because of a nearby brighter star).  For that star
we list Tycho-2 magnitudes \citep{hog00}.  $K$ magnitudes are taken from
the 2MASS point source catalog \citep{Cutri03}.}
\label{trendstars_table}
\end{deluxetable*}

%% file: asas_data_table_stub.tex
\begin{deluxetable}{lcc}
\tablecaption{ASAS Photometry of Boyajian's Star}
\tablewidth{0pt}
\tablehead{
\colhead{HJD} &
\colhead{$V$} &
\colhead{$\Delta V$} \\
\colhead{} &
\colhead{[mag]} &
\colhead{[mag]} 
}
\startdata
2453887.00345 & 11.928 & 0.019 \\
2453892.97057 & 11.922 & 0.034 \\
2453897.03348 & 11.937 & 0.035 \\
2453915.01509 & 11.923 & 0.028 \\
2453916.95754 & 11.940 & 0.037 \\
2453920.95091 & 11.955 & 0.035 \\
2453928.97449 & 11.941 & 0.031 \\
2453930.97710 & 11.922 & 0.017 \\
2453932.97819 & 11.918 & 0.027 \\
2453934.97840 & 11.932 & 0.057 
\enddata
\tablecomments{This table is available in its entirety in the electronic
edition of the journal.  A portion is reproduced here to provide
guidance on form and content.}
\label{asas_data_table}
\end{deluxetable}

%% file: photom_table.tex
\begin{deluxetable*}{lcccccccc}
\tablecaption{Photometry of Boyajian's Star}
\tablewidth{0pt}
\tablehead{
\colhead{Source} &
\colhead{$U$} &
\colhead{$B$} &
\colhead{$V$} & 
\colhead{$I$} & 
\colhead{$g$} & 
\colhead{$r$} & 
\colhead{$i$} &
\colhead{Observation} \\
\colhead{} &
\colhead{} &
\colhead{} & 
\colhead{} & 
\colhead{} & 
\colhead{} & 
\colhead{} & 
\colhead{} &
\colhead{Date}
}
\startdata
Tycho-2\tablenotemark{a} & ... & $12.80 \pm 0.25$\tablenotemark{b} & $12.02 \pm 0.22$\tablenotemark{b} & ... & ... & ... & ... & 1989-1993 \\
APASS\tablenotemark{c} & ... & $12.36 \pm 0.04$\tablenotemark{\phn} & $11.85 \pm 0.05$\tablenotemark{\phn} & ... & $12.05 \pm 0.05$\tablenotemark{\phn} & $11.70 \pm 0.06$\tablenotemark{\phn} & $11.55 \pm 0.04$\tablenotemark{\phn} & 2010-2012 \\
Kepler UBV Survey\tablenotemark{d} & $12.58 \pm 0.02$\tablenotemark{\phn} & $12.49 \pm 0.02$\tablenotemark{\phn} & $11.96 \pm 0.02$\tablenotemark{\phn} & ... & ... & ... & ... & 2011 \\
Kepler INT Survey & $12.68 \pm 0.11$\tablenotemark{e} & $12.57 \pm 0.09$\tablenotemark{e} & $11.75 \pm 0.07$\tablenotemark{e} & $10.91 \pm 0.11$\tablenotemark{e} & $12.17 \pm 0.06$\tablenotemark{f} & $11.46 \pm 0.12$\tablenotemark{f} & $11.47 \pm 0.10$\tablenotemark{f} & 2012 \\
WTF\tablenotemark{g} & ... & $12.26 \pm 0.01$\tablenotemark{\phn} & $11.71 \pm 0.02$\tablenotemark{\phn} & $11.05 \pm 0.10$\tablenotemark{\phn} & ... & ... & ... & 2014 \\
ASAS-SN & ... & ... & $11.91 \pm 0.01$\tablenotemark{\phn} & ... & ... & ... & ... & 2015-2017 \\
ASAS & ... & ... & $11.93 \pm 0.03$\tablenotemark{\phn} & $11.20 \pm 0.03$\tablenotemark{\phn} & ... & ... & ... & 2006-2017
\enddata
\label{photom_table}
\tablenotetext{a}{\citet{hog00}.}
\tablenotetext{b}{Converted from BT and VT magnitudes using the
  transformation equations given at https://heasarc.nasa.gov/W3Browse/all/tycho2.html.}
\tablenotetext{c}{\citet{Henden14}.}
\tablenotetext{d}{\citet{everett12}.}
\tablenotetext{e}{Converted from the SDSS magnitudes in this row using
the equations derived by \citet{jordi06}.  For $U$ and $B$
\citeauthor{jordi06} provide two equations, and we take the average
result of the two.}
\tablenotetext{f}{Converted from the INT/WFC Vega magnitudes reported
  by \citet{greiss12a,greiss12b} to SDSS AB magnitudes by inverting the
  transformation equations given by \citet{gs11}.  The uncertainties
  reported by \citet{greiss12b} appear to be unrealistically small
  (0.001~mag), so we adopt the rms of each magnitude as its uncertainty.}
\tablenotetext{g}{\citet{boyajian16}.}

\end{deluxetable*}